\documentclass[referee]{aa}
\usepackage{graphicx}
\usepackage{txfonts}
\usepackage{cases}

\begin{document}
   \title{Parametric survey of longitudinal prominence oscillation simulations}

   \author{Q. M. Zhang\inst{1,2} \and
           P. F. Chen\inst{1,3} \and
           C. Xia\inst{4} \and
           R. Keppens\inst{4} \and
           H. S. Ji\inst{2}
           }

   \institute{School of Astronomy and Space Science, Nanjing University,
              Nanjing 210093, China
              \and
              Key Laboratory for Dark Matter and Space Science, Purple
              Mountain Observatory, CAS, Nanjing 210008, China\\
		\email{zhangqm@pmo.ac.cn}
              \and
              Key Lab of Modern Astronomy and Astrophysics, Ministry of
              Education, China
              \and
              Centre for mathematical Plasma Astrophysics, Department of Mathematics,
              KU Leuven, Celestijnenlaan 200B, 3001 Heverlee, Belgium\\
              }

   \date{Received; accepted}
    \titlerunning{Simulating prominence oscillations}
    \authorrunning{Zhang et al.}

  \abstract
    {Longitudinal filament oscillations recently attracted more and more
     attention, while the restoring force and the damping mechanisms are still
     elusive.}
    {In this paper, we intend to investigate the underlying
    physics for coherent longitudinal oscillations of the entire
    filament body, including their triggering mechanism, dominant
    restoring force, and damping mechanisms.}
    {With the MPI-AMRVAC code, we carry out radiative hydrodynamic 
    numerical simulations of the longitudinal prominence oscillations. Two
    types of perturbations, i.e., impulsive heating at one leg of the loop 
    and an impulsive momentum deposition are introduced to the prominence, 
    which then starts to oscillate. We study the resulting oscillations for a
    large parameter scan, including the chromospheric heating duration,
    initial velocity of the prominence, and field line geometry.}
   {It is found that both microflare-sized impulsive heating at one leg of the
    loop and a suddenly imposed velocity perturbation can propel the prominence
    to oscillate along the magnetic dip. An extensive parameter survey results
    in a scaling law, showing that the period of the oscillation, which weakly
    depends on the length and height of the prominence, and the amplitude of the
    perturbations, scales with $\sqrt{R/g_\odot}$, where $R$ represents the
    curvature radius of the dip, and $g_\odot$ is the gravitational acceleration
    of the Sun. This is consistent with the linear theory of a pendulum, which
    implies that the field-aligned component of gravity is the main
    restoring force for the prominence longitudinal oscillations, as confirmed
    by the force analysis. However, the gas pressure gradient becomes 
    non-negligible for short prominences. The oscillation damps with time in
    the presence of non-adiabatic processes. Compared to heat conduction, 
    the radiative cooling is the dominant factor leading to the damping. A 
    scaling law for the damping timescale is derived, i.e., $\tau\sim l^{1.63}
    D^{0.66}w^{-1.21}v_{0}^{-0.30}$, showing strong dependence on the 
    prominence length $l$, the geometry of the magnetic dip (characterized by 
    the depth $D$ and the width $w$), and the velocity perturbation amplitude 
    $v_0$. The larger the amplitude, the faster the oscillation damps. It is
    also found that mass drainage significantly reduces the damping timescale
    when the perturbation is too strong.}
    {}


   \maketitle

\section{Introduction} \label{S-intro}

Solar prominences, or filaments when appearing on the solar disk, are cold and
dense plasmas suspended in the corona (Tandberg-Hanssen \cite{tan95}; Labrosse
et al. \cite{lab10}; Mackay et al. \cite{mac10}). They are formed above the
magnetic polarity inversion lines. The denser material is believed to be
supported by the magnetic tension force of the dip-shaped magnetic field lines
(Kippenhahn \& Schl{\"u}ter \cite{kip57}; Kuperus \& Raadu \cite{kup74}; Guo
et al. \cite{guo10}; Zhang et al. \cite{zhang12}; Xu et al. \cite{xu12}; Su \&
van Ballegooijen \cite{su12}). These fascinating phenomena attracted a lot of
simulation efforts from different aspects, such as their formation,
oscillations, and eruptions. With respect to the formation, the chromospheric
evaporation plus coronal condensation model has been studied widely with
one-dimensional (1D) simulations (e.g., M{\"u}ller et al. \cite{mul04};
Karpen et al. \cite{karp05,karp06}; Karpen \& Antiochos \cite{karp08}; Antolin
et al. \cite{antolin10}; Xia et al. \cite{xia11}; Luna et al. \cite{ln12b}),
where no back-reaction on the field topology is accounted for. It was then
for the first time extended to 2.5D by Xia et al. (\cite{xia12}) who simulated
the in situ formation of a filament in a sheared magnetic arcade and showed
that the condensation self-consistently forms magnetic dips while ensuring 
force-balance states. This finding strengthens the hitherto invariably 1D analysis 
performed for prominence formation and evolutions, as adopted by many authors 
to date. Once a prominence is formed, it might be triggered to deviate
from its equilibrium position and start to oscillate.

Observations demonstrate that prominences are hardly static. Besides 
small-amplitude oscillations (Okamoto et al. \cite{oka07}; Ning et al.
\cite{ning09}), large-amplitude and long-period prominence oscillations have 
been observed (e.g., Eto et al. \cite{eto02}; Isobe \& Tripathi 
\cite{iso06}; Gilbert et al. \cite{gil08}; Chen et al. \cite{chen08}; Tripathi
et al. \cite{tri09}; Hershaw et al. \cite{her11}; Bocchialini et al. 
\cite{bocc11}). The observations of the prominence oscillations led to the
comprehensive research topic of prominence seismology (Blokland \& Keppens
\cite{blok11a, blok11b}; Arregui \& Ballester \cite{arre11}; Arregui et al.
\cite{arre12}; Luna \& Karpen \cite{luna12}; Luna et al. \cite{ln12a}), 
and the long-term oscillations were considered as one of the precursors for
coronal mass ejections (CMEs; Chen, Innes, \& Solanki \cite{chen08}; Chen 
\cite{chen11}). Of particular interest in this paper are the longitudinal
oscillations along the axis of prominences/filaments, which were first 
presented in the simulation results of Antiochos et al. (\cite{anti00})
discovered from H$\alpha$ observations by Jing et al. (\cite{jing03}). The
phenomenon was further investigated by Jing et al. (\cite{jing06}) and
Vr{\v s}nak et al. (\cite{vrn07}). Such large-amplitude oscillations are
triggered by small-scale solar eruptions near the footpoints of the
main filaments, such as mini-filament eruptions, subflares, and flares. The
initial velocities of the oscillations are 30--100 km s$^{-1}$. The oscillation
period ranges from 40 min to 160 min and the damping times are $\sim$2--5 
times the oscillation period (Jing et al. \cite{jing06}).

Unlike the transverse oscillations whose restoring force is known to be the
magnetic tension force, the dominant restoring force for the longitudinal
oscillations still await to be clarified. Jing et al. (\cite{jing03}) proposed
several candidates for the restoring force, i.e., gravity, the pressure
imbalance, and the magnetic tension force. Vr{\v s}nak et al. (\cite{vrn07})
suggested that the restoring force is the magnetic pressure gradient along the
filament axis. With radiative hydrodynamic simulations, Luna \& Karpen 
(\cite{luna12}) and Zhang et al. (\cite{zhang12}) suggested that the gravity
component along the magnetic field is the main restoring force.
Li \& Zhang (\cite{li12}), on the other hand, suggested that both gravity and 
magnetic tension force contribute to the restoring force. As for the
damping mechanism, it really depends on the oscillation mode. For the vertical
oscillations, Hyder (\cite{hyde66}) proposed that the magnetic viscosity
contributes to the decay. For the horizontal transverse oscillations, Kleczek
\& Kuperus (\cite{kle69}) proposed that the induced compressional wave in the
surrounding corona acts to seemingly dissipate the oscillatory power. 
More damping mechanisms have been proposed, such as thermal conduction,
radiation, ion-neutral collisions, resonant absorption, and wave leakage
(see Arregui et al. \cite{arre12} and Tripathi et al. \cite{tri09} for
reviews). For the longitudinal oscillations, Zhang et al. (\cite{zhang12})
found that non-adiabatic terms such as the radiation and the heat conduction
contribute to the damping, but they might not be sufficient to explain the 
observed shorter timescale. In their simulations the chromospheric
heating is
switched off, so that the prominence mass was nearly fixed. On the contrary,
Luna \& Karpen (\cite{luna12}) studied the prominence oscillations while 
keeping the chromospheric heating and the resulting chromospheric evaporation.
As a result, the prominence was growing in length and mass during 
oscillations. They found that there are two damping timescales, a short one
for the initial stage and a longer one later. The analytical solution indicates
that the mass accumulation can explain the fast damping of the initial state.
As for the later slower damping, they suggested non-adiabatic effects such as
radiation and heat conduction. A quantitative survey is in order
to clarify how different geometrical and physical parameters of the prominence
affect the damping timescale.

Within the framework of gravity serving as the restoring force for the
filament longitudinal oscillations, in this paper we try to do a parameter
survey, aiming to clarify how the geometry of the magnetic field affects
the oscillation period and how the combined effects of radiation and heat 
conduction contribute to the damping of the oscillations. We describe the
numerical method in Section~\ref{S-method}. After showing the effects of the
perturbation type in Section~\ref{S-pertur}, we display the results of our
parameter survey in Section~\ref{S-results}. Discussions and summary are
presented in Sections~\ref{S-discussion} and \ref{S-summary}.

\section{Numerical method} \label{S-method}

High-resolution observations indicate that a filament/prominence is
made of many thin threads which are believed to be aligned to the individual
magnetic tubes (Lin et al. \cite{lin05}). Since the magnetic field inside the
filament is quite strong (Schmieder \& Aulanier \cite{schm12}) and the 
plasma beta is very low ($\beta\sim0.01-0.1$) (Antiochos et al.
\cite{anti00}; DeVore \& Antiochos \cite{dev00}; Aulanier et al. \cite{aul06}),
plus that the thermal conduction is strongly prevented across the field lines,
the dynamics inside different magnetic tubes can be considered to be
independent. Therefore, the formation and evolution of a filament thread 
can be treated as a 1D hydrodynamic problem.
Following Xia et al. (\cite{xia11}), the 1D radiative
hydrodynamic equations, shown as follows, are numerically solved by the
state-of-the-art MPI-Adaptive Mesh Refinement-Versatile Advection Code 
(MPI-AMRVAC; Keppens et al. \cite{kepp03,kepp12}).

\begin{equation} \label{eqn1}
\frac{\partial \rho}{\partial t}+\frac{\partial}{\partial s}(\rho v)=0 \,,
\end{equation}

\begin{equation} \label{eqn2}
\frac{\partial}{\partial t}(\rho v)+\frac{\partial}{\partial s}(\rho v^2+p)=
\rho g_{\parallel}(s) \,,
\end{equation}

\begin{equation} \label{eqn3}
\frac{\partial \varepsilon}{\partial t}+\frac{\partial}{\partial s}
(\varepsilon v+pv)=\rho g_{\parallel}v+H-n_{\rm H}n_{\rm e}\Lambda(T)+
\frac{\partial}{\partial s}(\kappa \frac{\partial T}{\partial s}) \,,
\end{equation}

\noindent
where $\rho$ is the mass density, $T$ is the temperature, $s$ is the
distance along the loop, $v$ is the velocity of plasma, $p$ is the gas
pressure, $\varepsilon=\rho v^2 /2 + p/(\gamma -1)$ is the total energy
density, $n_{\rm H}$ is the number density of hydrogen, $n_{\rm e}$ is
the number density of electrons, and $g_\parallel(s)$ is the component
of gravity at a distance $s$ along the magnetic loop, which is determined
by the geometry of the magnetic loop. Furthermore,
$\gamma=5/3$ is the ratio of the specific heats, $\Lambda(T)$ is the
radiative loss coefficient for the optically thin emission, $H(s)$ is
the volumetric heating rate, and $\kappa=10^{-6} T^{5/2}$ ergs cm$^{-1}$
s$^{-1}$ K$^{-1}$ is the Spitzer heat conductivity. As done in
previous works mentioned in \S\ref{S-intro}, we assume a fully ionized
plasma and adopt the one-fluid model. Considering the helium abundance
($n_{\rm He}/n_{\rm H}=0.1$), we take $\rho=1.4 m_{\rm p} n_{\rm H}$ and
$p=2.3 n_{\rm H} k_{\rm B} T$, where $m_{\rm p}$ is the proton mass and 
$k_{\rm B}$ is the Boltzmann constant. 
Note that the above equations are different from those in Luna \&
Karpen (\cite{luna12}) in that a uniform cross section is assumed here for the
flux tube for simplicity, where expanding flux tubes based on given, immobile 3D magnetic
fields are adopted in Luna \& Karpen (\cite{luna12}).
The radiative hydrodynamic equations 
(\ref{eqn1}--\ref{eqn3}) are numerically solved by the MPI-AMRVAC code, 
where the heat conduction term is solved with an implicit scheme separately
from other terms. To include the radiative loss, we take the
second-order polynomial interpolation to compile a high resolution table 
based on the radiative loss calculations using updated element abundances
and better atomic models over a wide temperature range (Colgan et al. 
\cite{colg08}). The corresponding values in this table are systematically 
$\sim$2 times larger than the previous radiative loss
function adopted by Luna \& Karpen (\cite{luna12}).

\begin{figure}
\includegraphics[width=10cm,clip=]{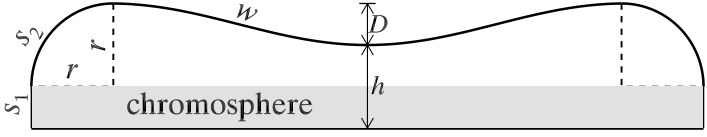}
\caption{Magnetic loop used for the 1D radiative hydrodynamic
simulations of the prominence oscillations. Note that the horizontal 
and the vertical sizes are not to scale.}
\label{fig1}
\end{figure}

It is often believed that a prominence is hosted at the dip of a magnetic
loop, supported by the magnetic tension force. Therefore, we adopt a loop
geometry with a magnetic dip, which is symmetric about the midpoint, as
shown in Fig. \ref{fig1}. The loop consists of two vertical legs with a
length of $s_1$, two quarter-circular shoulders with a radius $r$ (the length
of each arc, $s_2-s_1$, is $\pi r/2$), and a quasi-sinusoidal-shaped dip with 
a half-length of $w$. The height of the dip is expressed as 
$y=D-D\cos(\pi x/2w)$
if the local coordinates ($x$, $y$) are centered at the midpoint of the dip.
The dip has a depth of $D$ below the apex of the loop. Such a geometry
determines the field-aligned component of the gravity, whose distribution along
the left half of the magnetic loop is expressed as follows:

\begin{equation} \label{eqn4}
g_\parallel(s)=\cases{-g_\odot , & $s \leqslant s_1$; \cr
-g_\odot \cos\left(\frac{\displaystyle \pi}{\displaystyle 2}\frac
{\displaystyle s-s_1}{\displaystyle s_2-s_1}\right), &
$s_1 < s \leqslant s_2$; \cr
g_\odot \frac{\displaystyle \pi D}{\displaystyle 2(L/2-s_2)}
\sin\left(\pi \frac{\displaystyle s-s_2}
{\displaystyle L/2-s_2}\right), & $s_2 < s \leqslant L/2$, \cr}
\end{equation}

\noindent
where the gravity at the solar surface $g_{\odot}=2.7\times 10^2$ m s$^{-2}$,
the total length of the loop $L$, the length of each vertical segment
$s_1=5$ Mm, and $s_2=s_1+\pi r/2$ Mm. The total length of the dip is 
$2w=L-2s_2$. The field-aligned component of the gravity in the right half is
symmetric to the left half. The parameter $h=s_1+r-D$ gives the height of the 
central dip above the lower boundary.

Our simulations start from a thermal and force-balanced equilibrium state where
the background heating is balanced by radiative loss and thermal conduction,
and the plasma in the loop is quiescent. The simulations are divided into three
steps: (1) Prominence formation: A prominence forms and grows near the center
of the magnetic dip as chromospheric material is evaporated into the corona and
condensates due to thermal instability after chromospheric heating is
introduced near the footpoints of the loop; (2) Prominence relaxation: The
prominence relaxes to a thermal and force-balanced equilibrium state as the
localized heating is halted and the chromospheric evaporation ceases; (3)
Prominence oscillation subjected to perturbations: The prominence starts to
oscillate with a damping amplitude after perturbations are introduced. In step
1, which lasts for a time interval of $\Delta t_1$, the heating term $H(s)$ in 
Eq.~(\ref{eqn3}) is composed of two terms, i.e., the steady background heating
$H_0(s)$ and the localized chromospheric heating $H_1(s)$, which are
expressed as follows:

\begin{equation} \label{eqn5}
H_0(s)=\cases{E_0 \exp(-s/H_m), & $s < L/2$; \cr
        E_0 \exp[-(L-s)/H_m], & $L/2 \leqslant s < L$; \cr}
\end{equation}

\begin{equation} \label{eqn6}
H_1(s)=\cases{E_1, & $s \leqslant s_{tr}$; \cr
E_1 \exp[-(s-s_{tr})/\lambda], & $s_{tr} < s \leqslant L/2$; \cr
E_1 \exp[-(L-s_{tr}-s)/\lambda], & $L/2 < s \leqslant L-s_{tr}$; \cr
E_1, & $s > L-s_{tr}$; \cr}
\end{equation}

\noindent
where the quiescent heating term $H_0$ is adopted to maintain the hot
corona with the amplitude $E_0=3\times 10^{-4}$ ergs cm$^{-3}$ s$^{-1}$ and
the scale-height $H_m=L/2$, and the localized heating term $H_1$ is
adopted to generate chromospheric evaporation into the corona with the
amplitude $E_1=10^{-2}$ ergs cm$^{-3}$ s$^{-1}$, the transition region height
$s_{tr}=6$ Mm, and the scale height $\lambda=10$ Mm. The heating is
taken to be symmetric in order to form a static prominence near the magnetic 
dip center, so that we can easily control the manner how the prominence is
triggered to oscillate. Our methodology is different from Luna \& Karpen
(\cite{luna12}) who used asymmetric heating which spontaneously leads to the
oscillation once the prominence is formed. In step 2, $H_1$ is 
switched off. Owing to the absence of the chromospheric evaporation, the gas 
pressure inside the magnetic loop drops down, so the compressed prominence 
expands until a new equilibrium is reached, which roughly takes less than 2.4 hr. 
In step 3, a perturbation is introduced to the prominence in order to trigger 
its oscillation. Note that $H_0$ remains throughout the simulations.

From the observational point of view, there might be two kinds of
perturbations. The first one is an impulsive momentum injected to the magnetic
loop as the magnetic reconnection near the footpoints rearranges the magnetic
loop rapidly. The second is impulsive heating due to subflares (e.g., Jing et
al. \cite{jing03}, Vr{\v s}nak et al. \cite{vrn07}, Li \& Zhang \cite{li12})
or microflares (Fang et al. \cite{fang06}) near the footpoints of the magnetic
loop where a large amount of magnetic energy is impulsively released through
magnetic reconnection. The gas pressure is greatly increased that could propel
the prominence to oscillate along the dip-shaped field lines. In our 1D
simulations, we separate the two effects to see their difference. 
In one case, a velocity perturbation with the following distribution
is imposed to the prominence,
 
\begin{equation} \label{eqn7}
v(s)=\cases{0, & $s < s_{\rm pl}-\delta$; \cr
v_0(s-s_{\rm pl}+\delta)/\delta, & $s_{\rm pl}-\delta \leqslant s \leqslant 
	s_{\rm pl}$; \cr
v_0, & $s_{\rm pl} \leqslant s \leqslant s_{\rm pr}$; \cr
v_0(-s+s_{\rm pr}+\delta)/\delta, & $s_{\rm pr} \leqslant s \leqslant 
	s_{\rm 	pr}+\delta$; \cr
0, & $s > s_{\rm pr}$, \cr}
\end{equation}
\noindent
where $s_{\rm pl}$ and $s_{\rm pr}$ are the coordinates of the left
and right
boundaries of the prominence, $\delta=10$ is the buffer zone which allows that
the perturbation velocity varies smoothly in space, and $v_0$ is the 
perturbation amplitude. In the other case, impulsive heating ($H_2$), as
described as follows, is introduced near the right-hand footpoint of the magnetic loop,

\begin{equation} \label{eqn8}
H_2(s)=E_2 \exp{\left[{-\frac{(s-s_{peak})^2}{s_{scale}^2}-
\frac{(t-t_{peak})^2}{t_{scale}^2}}\right]},
\end{equation}
\noindent
where the heating spatial scale $s_{scale}=2.5$ Mm, the peak location
$s_{peak}=245$ Mm, the heating timescale $t_{scale}=5$ min, and the peak time
$t_{peak}=15$ min. The heating ramps up to the peak for 15 min and then fades
down to 0.

As for the boundary conditions, all variables
at the two footpoints of the magnetic loop are fixed, which is justified 
because the density in the low atmosphere is more than four orders of 
magnitude higher than that in the corona. The same approach has been adopted 
by Ofman \& Wang (\cite{ofm02}) and Xia et al. (\cite{xia11}), assuming that 
the coronal dynamics has little effect on the low atmosphere. The approach
was verified by Hood (\cite{hood86}) with the parameters being far from the
marginal stability. The violation of the rigid wall conditions in certain
cases was discussed by van der Linden et al. (\cite{van94}).

\section{Effects of the perturbation type} \label{S-pertur}

In order to check how the two types of perturbations as described in 
\S\ref{S-method} influence the characteristics of the prominence oscillations,
we perform simulations of the oscillations which are excited by the two types
of perturbations while keeping $\Delta t_1=7.2$ hr, $r=20$ Mm, $D=10$ Mm, and 
$L=260$ Mm.

\begin{figure}
 \centerline{\hspace*{0.015\textwidth}
             \includegraphics[width=0.45\textwidth,clip=]{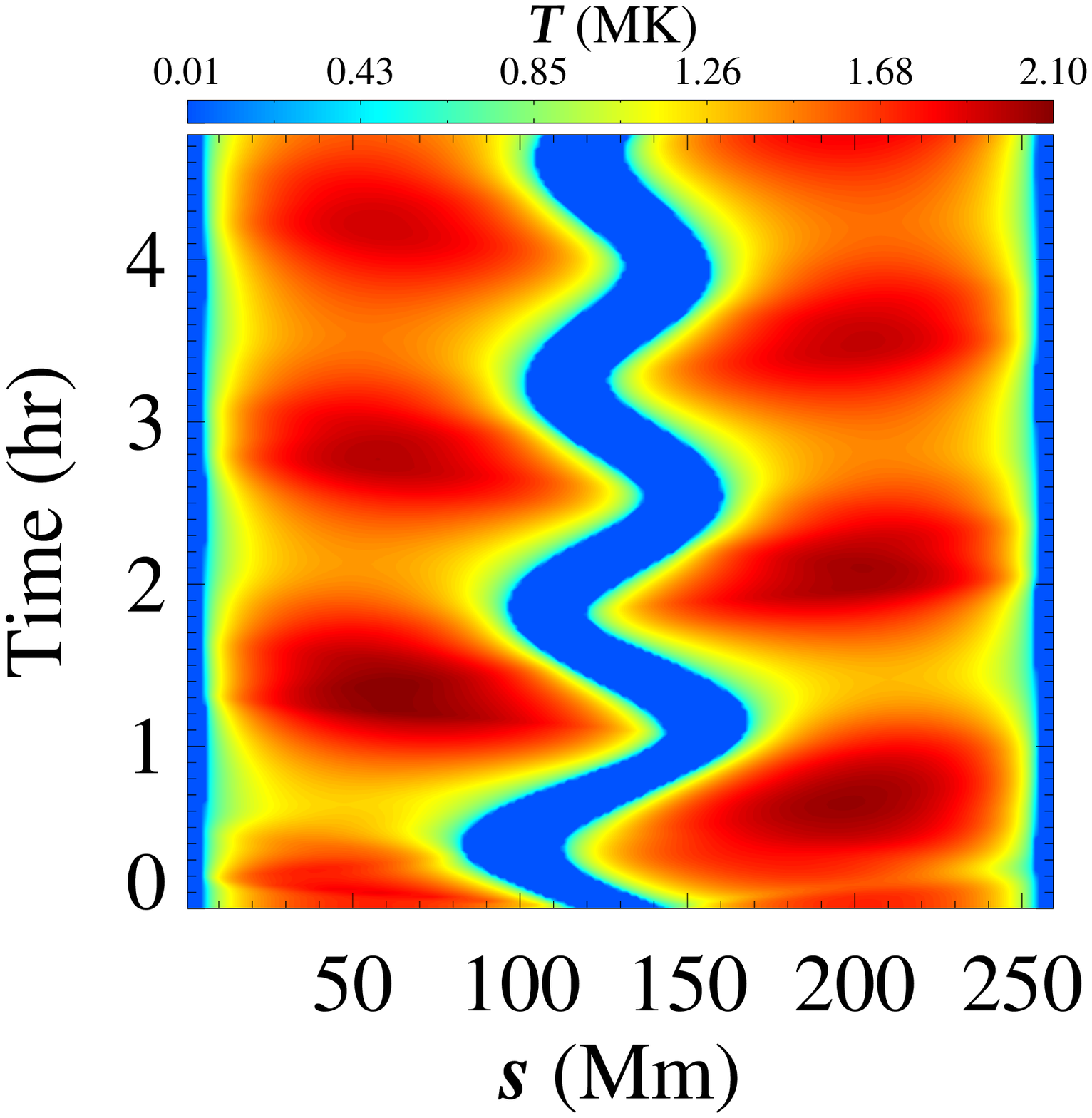}
             \hspace*{0.0\textwidth}
             \includegraphics[width=0.45\textwidth,clip=]{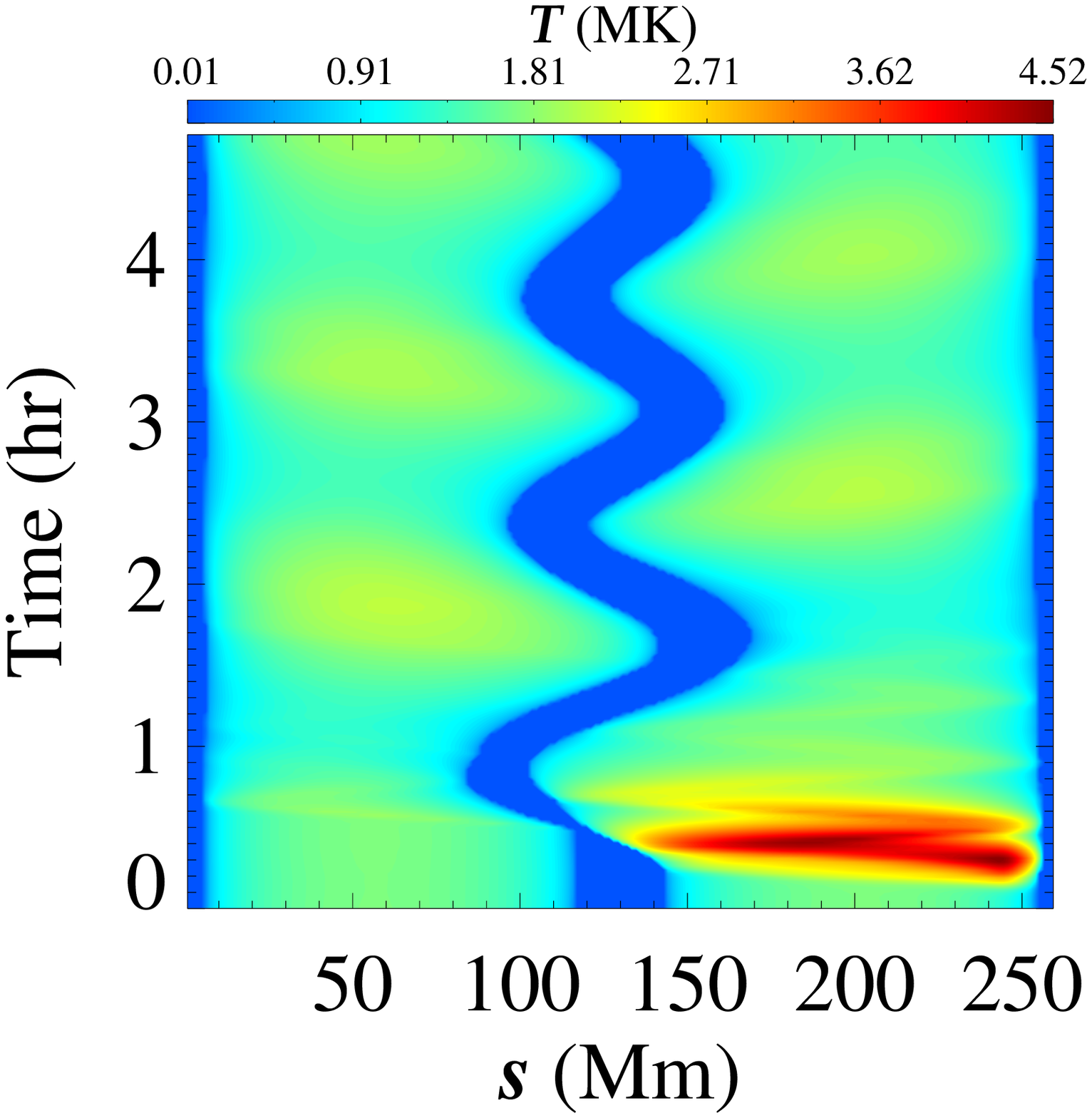}
            }
\caption{Comparison of the evolutions of the temperature of the loop 
between the two types of perturbations. The left panel corresponds to the case
with velocity perturbations with $v_0=-40$ km s$^{-1}$ and the right panel to
the case with localized heating perturbations with $E_2=0.24$ 
ergs cm$^{-3}$ s$^{-1}$.}
\label{fig2}
\end{figure}

In case A, the prominence oscillation is triggered by a velocity perturbation
over the whole prominence body. With $v_0=-40$ km s$^{-1}$ (the minus means
that the velocity is toward the left), the temporal evolution of the plasma
temperature distribution along the magnetic loop is displayed in the left
panel of Fig. \ref{fig2}. It is seen that in response to the perturbation, the
prominence, signified by the low temperature, starts to oscillate around the
equilibrium position. The oscillation amplitude decays with time. Fitting the
trajectory of the mass center of the oscillating prominence with a decayed 
sine function

\begin{equation}\label{eq-sin}
s=s_0+A_0\sin({2\pi \over P}t+\phi_0)\exp{(-t/\tau)},
\end{equation}
\noindent
we find the initial amplitude $A_0=34.9$ Mm, the oscillation period $P=84.3$
min, and the damping timescale $\tau=272$ min. Assuming that the prominence
thread has a cross-section area of $\sim 3.14\times 10^{14}$ cm$^2$ (Lin et
al. \cite{lin05}), the initial kinetic energy of the oscillating prominence
thread is estimated to be $\sim7.2\times 10^{23}$ ergs. It is noted
that the single decayed sine function, as used for fitting the H$\alpha$ 
observations (Jing et al. \cite{jing03}; Vr{\v s}nak et al. \cite{vrn07};
Zhang et al. \cite{zhang12}), fits the simulated observations very well. On
the contrary, a combination of Bessel function and an exponential decay
function is necessary to fit the initial overtone in the simulations of Luna
\& Karpen (\cite{luna12}), which results from the continual mass accumulation.

In case B, the prominence oscillation is triggered by the impulsive heating
which is deposited near the right leg of the magnetic loop in order to mimic
a microflare near the prominence. To do that, an impulsive heating term
$H_2(s)$ in Eq.~(\ref{eqn8}) is added to the heating term $H$ in Eq.
(\ref{eqn3}), where $s_{peak}=245$ Mm meaning the heating is concentrated at a
height of 15 Mm above the right footpoint of the magnetic loop. 

The right panel of Fig. \ref{fig2} depicts the 
temporal evolution of the temperature distribution along the magnetic 
loop with $E_2=0.24$ ergs cm$^{-3}$ s$^{-1}$.
With the typical cross-section area of a prominence thread being 
$\sim$3.14$\times$10$^{14}$ cm$^2$, the corresponding total energy deposited
into the single magnetic loop $E_{heating}$ is 1.8$\times 10^{25}$ ergs.
This value is reasonable since observations indicate that the total 
energy of a microflare is 10$^{26}$--10$^{27}$ ergs or even more (e.g., Shimizu
et al. \cite{shim02}; Hannah et al. \cite{han08}; Fang et al. \cite{fang10}),
and several percent of the released energy goes into one prominence thread.
From another point of view, under the framework of magnetic reconnection model
for microflares, the magnetic energy release rate is estimated to be
$B^{2}v_{in}/(4\pi L)$. With the magnetic field $B\sim 20$ G, the reconnection
inflow speed $v_{in}$ being about 0.1 times the Alfv\'en speed which is
about 1000 km s$^{-1}$ (Jiang et al. \cite{jia12}), and the spatial size $L=
10\arcsec$, the energy release rate is estimated to be $\sim 0.88$ ergs
cm$^{-3}$ s$^{-1}$, which is of the order adopted here. Fitting the trajectory 
of the oscillating prominence
with the damped sine function as shown in Eq. (\ref{eq-sin}) yields $A_0=35.8$
Mm, $P=84.3$ min, and $\tau=268$ min. The corresponding initial velocity is also 
-40 km s$^{-1}$. This indicates that a typical microflare near the leg of
the magnetic loop hosting a prominence thread can excite the prominence
longitudinal oscillations with an initial velocity of tens of km s$^{-1}$. The
corresponding kinetic energy is only $\sim7.2\times 10^{23}/1.8\times 
10^{25}$, i.e., $\sim$4\% of the deposited thermal energy.
The remaining $\sim$96\% of the energy deposit contributes to the
heating of the chromosphere.

\section{Parameter survey} \label{S-results}

The results in \S\ref{S-pertur} reveal that the oscillation period
does not strongly depend on the two types of perturbations, i.e., impulsive momentum
and localized heating at one footpoint used in our investigation. 
Note that we concentrate on the oscillation characteristics which follow the 
small transient/excitation phase, already obtained from simple decaying sinusoidal 
fitting. A small difference in the decay timescale
exists between the two perturbation types. With the same initial velocity, the
decay timescale is 4 minutes shorter in the case of impulsive heating than that
in the case of impulsive momentum. However, the relative variation, 1.4\%, is
very small. Therefore, we can conclude that the oscillation is basically
intrinsic and the characteristics of the oscillation depend on the prominence
itself and the geometry of the magnetic loop in our case where there
is no mass accumulation and the oscillations are excited by either impulsive
momentum or localized heating. The prominence feature is only 
characterized by the thread length ($l$), and the geometry of the magnetic
loop is characterized by $r$, $D$, and $w$ as depicted in Fig. \ref{fig1}.
Among the three geometrical parameters, $h=s_1+r-D$ determines the height of
the prominence, $D$ and $w$ determine the curvature of the magnetic dip. If
other parameters are fixed, the length of the prominence is determined by the
duration of the chromospheric evaporation in step 1, i.e., $\Delta t_1$, as 
described in \S\ref{S-method}. Besides, the decay timescale might vary
with the perturbation amplitude, therefore another parameter is the initial
perturbation velocity $v_0$. In this section, we perform a parameter survey
to investigate how each individual one among the five parameters 
($\Delta t_1$, $r$, $D$, $w$, and $v_0$) changes the oscillation period and
the decay timescale. For each parameter, several cases with different values
are simulated with other parameters fixed. In our simulations, 
we set $r=10$ Mm, $D=5$ Mm, $w=110$ Mm, and $v_0=-20$ km s$^{-1}$
  when varying $\Delta t_1$. 
We set $\Delta t_1=7.16$ hr, $D=5$ Mm, $w=90$ Mm, and $v_0=-20$ km s$^{-1}$   
  when varying $r$. 
We set $\Delta t_1=7.16$ hr, $D=5$ Mm, $r=10$ Mm, and $v_0=-20$ km s$^{-1}$
  when varying $w$.
We set $\Delta t_1=7.16$ hr, $r=20$ Mm, $w=93.6$ Mm, and $v_0=-20$ km s$^{-1}$
  when varying $D$. 
We set $\Delta t_1=7.16$ hr, $r=20$ Mm, $w=93.6$ Mm, and $D=10$ Mm
  when varying $v_0$.
Since the oscillation characteristics are found nearly insensitive 
to the perturbation type, we use the velocity perturbation to excite the 
oscillations in the survey.

\subsection{Length and mass of the prominence} \label{S-length}

\begin{figure}
\includegraphics[width=10cm,clip=]{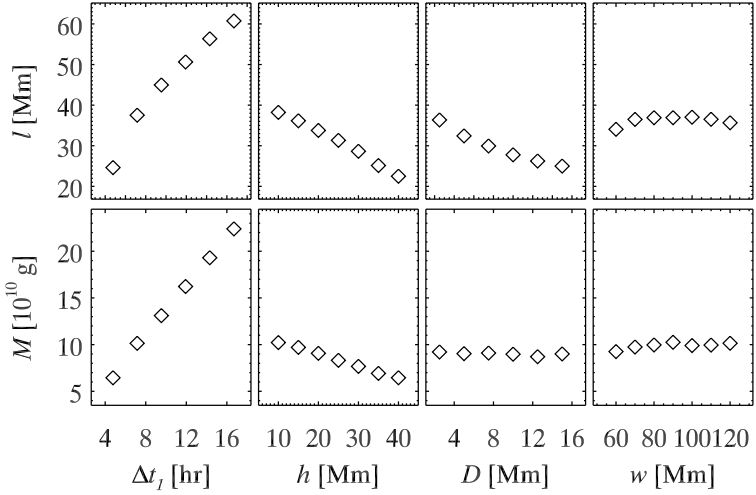}
\caption{Scatter-plots of the total length $l$ (upper panels) and mass $M$
(lower panels) of the prominences at the end of relaxation step as functions 
of $\Delta t_1$, $h$, $D$, and $w$.}
\label{fig3}
\end{figure}

After finishing the first two steps of the simulations as described in 
\S\ref{S-method}, we get a quasi-static prominence. The dependence of the
prominence length $l$ on $\Delta t_1$, $h$, $D$, and $w$ is shown in the four
panels of the upper row of Fig. \ref{fig3}. It can be seen that $l$, which
fits into the scaling law $l\sim \Delta t_1^{0.70}$, increases with the
duration of the heating time $\Delta t_1$. It is understandable since more
chromospheric plasma is evaporated into the corona when $\Delta t_1$ increases.
The length $l$ decreases with $h$ as $l\sim h^{-0.37}$, which is probably
because it takes a longer time for the more tenuous corona to condensate as
the height of the magnetic dip increases, and therefore the effective heating
time is shorter. The length $l$ decreases with $D$ as $l\sim D^{-0.21}$,
which can be understood as the prominence becomes more compressed as the
magnetic dip becomes deeper. However, the length of the prominence does not
vary considerably with $w$. Of course, $w$ should not be too small, otherwise
thermal instability would not occur. The lengths of these simulated prominence
threads are consistent with the reported values, i.e., tens of Mm (Lin et al.
\cite{lin05}).

The dependence of the prominence mass $M$ on $\Delta t_1$, $h$, $D$, and $w$ 
is shown in the four panels of the lower row of Fig. \ref{fig3}. It can be 
seen that the dependence of $M$ on $\Delta t_1$, $h$, and $w$ is similar to
$l$. Their difference is that $l$ decreases with $D$ whereas $M$ does not
change with $D$, which means that the plasma number density 
(10$^{10}$--10$^{11}$ cm$^{-3}$, and the corresponding density is
10$^{-14}$--10$^{-13}$ g cm$^{-3}$) is higher in the prominence 
with a deeper magnetic dip. A scaling law is obtained by fitting
the data points, which is $M\sim \Delta t_1^{0.98}h^{-0.34}$.

It is noted that the above results are derived for a dipped magnetic loop
filled via chromospheric evaporation with a limited lifetime, where the 
prominence thread can sustain in the corona. In the case of magnetic loops
without a dip (e.g., Mendoza-Brice{\~n}o et al. \cite{men05}) or with a 
shallow dip and asymmetric heating (e.g., Karpen et al. \cite{karp06}),
condensations repetitively form, stream along the magnetic field, and
ultimately disappear after falling back to the nearest footpoint. Therefore,
the mass and length of the prominence evolve dynamically, without reaching
an equilibrium value.

\subsection{The oscillation period and decay timescale} \label{S-period}

As the velocity perturbation is introduced to the quasi-static prominence, the
prominence starts to oscillate. Fitting the trajectory of the oscillating
prominence with the damped sine function shown in Eq. (\ref{eq-sin}), we get
the oscillation period ($P$) and the decay timescale ($\tau$) for each case 
in the parameter survey.

\begin{figure}
\includegraphics[width=11cm]{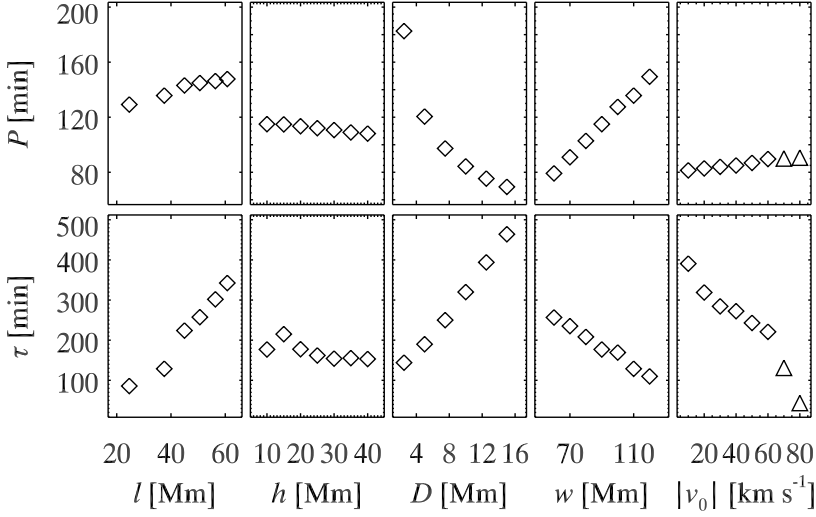}
\caption{Scatter-plots of the period $P$ (upper panels) and damping time $\tau$
(lower panels) of the prominences in the oscillation step as functions of
$l$, $h$, $D$, $w$, and $v_0$. The values of $P$ and $\tau$ in the cases
$|v_0|=70$ and 80 km s$^{-1}$ that cause mass drainage at the footpoint of
the coronal loop are marked with triangles in the right panels.}
\label{fig4}
\end{figure}

The variations of $P$ along with the parameters $l$, $h$, $D$, $w$, and $v_0$ 
are shown in the upper row
of Fig. \ref{fig4}. It is seen that $P$ increases slightly with $l$
and $v_0$, and decreases slightly with $h$. However, it increases seriously with
$w$ and decreases with $D$. To fit the variations with a scaling
law, we obtain $P\sim l^{0.16}h^{-0.05}D^{-0.54}w^{0.91}v_{0}^{0.05}$. 
Therefore, the period of prominence longitudinal
oscillations relies dominantly on the geometry of the dip, especially its
curvature. It is noted that the range of $P$ is in agreement with the reported 
values in previous studies (e.g., Jing et al. \cite{jing06}).

The variations of $\tau$ along with the five parameters are shown in the lower
row of Fig. \ref{fig4}. It is seen that $\tau$ increases significantly with
$l$ and $D$, and decreases with $w$ and $v_0$. It is noted that in the cases
of $|v_0|=70$ and 80 km s$^{-1}$, part of the prominence mass drains down to
the chromosphere, which is why the triangles in the lower-right panel of 
Fig.~\ref{fig4} do not follow the trend of the data points denoted by the
diamonds where $|v_0| < 70$ km s$^{-1}$. The decay timescale does not vary
significantly with $h$. To fit the variations with a scaling law, we obtain 
$\tau\sim l^{1.63}h^{-0.18}D^{0.66}w^{-1.21}v_{0}^{-0.30}$, where the
cases with prominence drainage are not included in the fitting. The values of
$\tau$ are also in the same order of magnitude as the observed ones.

\section{Discussions} \label{S-discussion}

\subsection{Restoring force} \label{S-force}

\begin{figure}
\includegraphics[width=10cm]{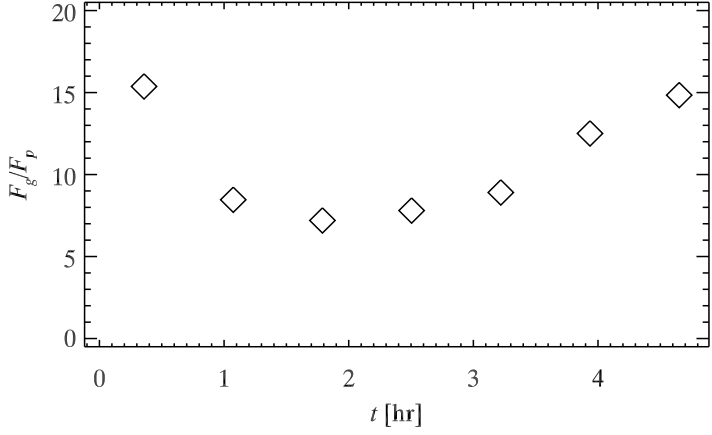}
\caption{Temporal variation of $F_{g}/F_{p}$ when the displacement of the prominence 
reaches maximum during each half-cycle in the case of $r=20$ Mm and
$D=10$ Mm. The velocity perturbation is -40 km s$^{-1}$.}
\label{fig5}
\end{figure}

For an oscillating phenomenon, the most important thing is the determination
of the restoring force, which directly decides the oscillation period. In our
1D hydrodynamic simulations, the only forces exerted on the prominence are 
the gravity and the gas pressure gradient, both are restoring forces for the
longitudinal oscillations. In order to compare their importance,
we calculate the two forces in the case with $\Delta t_1=7.16$ hr, $v_0=-40$ km
s$^{-1}$, $r=20$ Mm, $D=10$ Mm, and $w=93.6$ Mm. The two forces are calculated
when the prominence is the furthest from the equilibrium position.
Despite that the plasma in prominences is hundreds of times denser than
the ambient corona, it is not an ideal rigid body. For oscillations with higher
modes as studied by Luna et al. (\cite{ln12a}), the pressure gradient changes
rapidly along the prominence thread. For the fundamental-mode oscillations in
this paper, the prominence oscillates as a whole and the pressure gradient
changes slightly along the thread. Therefore, for simplicity, we compare the
overall magnitude of the two forces by a simple calculation instead of as
point-to-point one in the simulations. The integral
of the gravity force is quantified between the two ends of the prominence, i.e., 
$F_g=\int_{left}^{right}\rho|g_\parallel|ds=\int_{left}^{right}\rho g_{\odot}
\frac{\pi D}{2w}|\sin(\frac{\pi(s-L/2)}{w})|ds$, where a unit area is assumed
for the cross section. The integral of pressure gradient force over the
prominence is expressed as 
$F_p=\int_{left}^{right}|\partial p/\partial s|ds=|p_{right}-p_{left}|$. 
The left and right boundaries of the prominence
are defined to be where the density drops to 7$\times 10^{-14}$ g cm$^{-3}$.
Figure \ref{fig5} displays the temporal evolution of the ratio $F_{g}/F_{p}$, 
from which it is seen that the gravitational force is generally $\sim$10 times
larger than the gas pressure gradient force.

Since the gravity is the dominant restoring force, the overall motion of the
prominence can also be described for simplicity as

\begin{equation} \label{eqn-newton}
M\frac{d^2x}{dt^2}=M g_{\parallel}=-M g_{\odot}\frac{\pi D}{2w}\sin(\frac{\pi x}{w}),
\end{equation}
\noindent
where $x=s-L/2$ is the displacement of the prominence from the equilibrium
position. It is not easy to solve this equation analytically. 
However, if the oscillation
amplitude is much smaller than the half width of the whole magnetic dip ($w$),
we get the approximation $\sin(\pi x/w)\approx \pi x/w$. So,
the above equation is simplified to be

\begin{equation} \label{eqn-new1}
M\frac{d^2x}{dt^2}=-M g_{\odot}\frac{\pi D}{2w}\frac{\pi x}{w},
\end{equation}
\noindent
with the solution $x=A_0\sin(\frac{2\pi}{P}t+\phi)$. The corresponding period is

\begin{equation} \label{eqn-new2}
P=\sqrt{\frac{8w^2}{g_{\odot}D}}.
\end{equation}

\begin{figure}
\includegraphics[width=10cm]{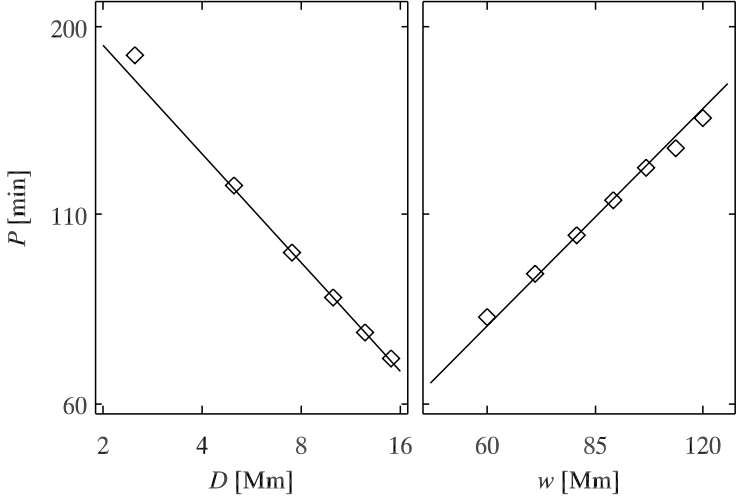}
\caption{Comparison of the periods of the prominence oscillations from
simulations ({\it diamonds}) and theoretical analysis ({\it solid line})
as a function of the depth of the magnetic dip $D$ ({\it left panel})
and the width of the dip $w$ ({\it right panel}). Note that both axes are
in logarithmic scale.}
\label{fig6}
\end{figure}

Such a period can also be readily obtained if the prominence is taken in analogy
to a pendulum whose period is 

\begin{equation} \label{eqn-new3}
P=2\pi \sqrt{\frac{R}{g_{\odot}}},
\end{equation}
\noindent
where $R$ is the curvature radius of the dipped magnetic loop. With the shape
of the loop being $y=D-D\cos(\pi x/2w)$, the curvature radius at the loop 
center is approximated to be $R=2w^2/(D\pi^2)$. Substituting $R$ into 
Eq. (\ref{eqn-new3}), we get $P=\sqrt{8w^2/(g_{\odot}D)}$, the same as 
Eq. (\ref{eqn-new2}). Figure \ref{fig6} compares the oscillation periods
obtained from the hydrodynamic simulations ({\it diamonds}) and those
estimated from Eq. (\ref{eqn-new2}) ({\it solid line}) when the two
parameters, $D$ and
$w$, are changed. It is revealed that Eq. (\ref{eqn-new2}) is a very good
approximation for estimating the period of the prominence longitudinal
oscillation. Of course, it should be kept in mind that the derivation of Eq.
(\ref{eqn-new2}) is based on the assumption that the dipped magnetic loop has
a sinusoidal shape. More generally, the oscillation period is related to the
local curvature radius $R$ by the formula $P=2\pi \sqrt{R/g_{\odot}}$, as also
demonstrated by Luna \& Karpen (\cite{luna12}).

Recently, Luna et al. (\cite{ln12a}) extended the theoretical analysis
of longitudinal prominence oscillations by including the effect of the pressure
gradient force. They found that the ultimate fundamental frequency of the
oscillations is found from $\omega_{\rm fund}^2=\omega_{g}^2+\omega_{s}^2$, where 
$\omega_{g}$ and $\omega_{s}$ stand for the gravity-driven and pressure-driven
frequencies, respectively. The ratio of the two frequencies 
$\omega_{g}^2/\omega_{s}^2=R_{\rm lim}/R$, where $R_{\rm lim}$ denotes the
critical value of the curvature radius ($R$) of the magnetic dip. If 
$R \ll R_{lim}$, then gravity dominates over pressure in the restoring force
of longitudinal oscillations. They pointed out that the reported values of the
curvature are small compared with $R_{\rm lim}$, so that it is reasonable to
ignore the effect of the pressure term in most cases. In our parameter survey,
$R_{\rm lim}=0.175(L-l)l$ ranges from 760 to 2100 Mm and the ratio $R/R_{\rm lim}$ ranges from 0.1 to 0.5. Hence, their theoretical results of 
gravity being the main restoring force for the fundamental mode 
in this parameter range are thus confirmed by our simulations.

For a prominence above the solar limb, all the parameters in Eq. 
(\ref{eqn-new2}) can be roughly measured. Combined with the results in this
paper, the comparison between simulations and observations in Zhang et al.
(\cite{zhang12}) implies that Eq. (\ref{eqn-new2}) is a good approximation
to estimate the oscillation period. For the prominence longitudinal
oscillations on the solar disk, i.e., filament longitudinal oscillations, only
the oscillation period can be unambiguously measured. Eq. (\ref{eqn-new2}) then
provides a diagnostic tool for inferring the geometry of the dipped magnetic
loop. Especially, when $w$ can be roughly estimated from force-free
magnetic extrapolations, the depth of the dip, $D$, can be determined. At
least, we can estimate the curvature radius of the dipped magnetic field, $R$,
through Eq. (\ref{eqn-new3}). After the determination of $R$, Luna \& Karpen
(\cite{luna12}) further proposed an approximate method to estimate the magnetic
field in the prominence.

Besides the dominant dependence on the geometric parameters, the oscillation
period also weakly changes with the length and the height of the prominence,
as well as with the initial velocity. These can be understood as follows: (1)
Dependence on the prominence length: As the prominence thread is shorter, the
ratio of the gas pressure gradient to the gravity would increase as indicated
by our simulations, therefore, the gas pressure gradient would contribute to
the restoring force, resulting in a shorter oscillation period; (2) Dependence
on the prominence height: As seen from Fig. \ref{fig3}, with other parameters
the same, a high prominence has a shorter length. Therefore, with the same
reason as in (1), the oscillation period would be smaller; (3) Dependence on
the initial velocity: Since $\sin(\pi x/w)$ is always smaller than
$\pi x/w$ in Eq. (\ref{eqn-newton}), the nonlinear term would naturally
lead to a long period as the oscillation amplitude increases.

\subsection{Damping mechanisms} \label{S-damp}

\begin{figure}
\includegraphics[width=8cm]{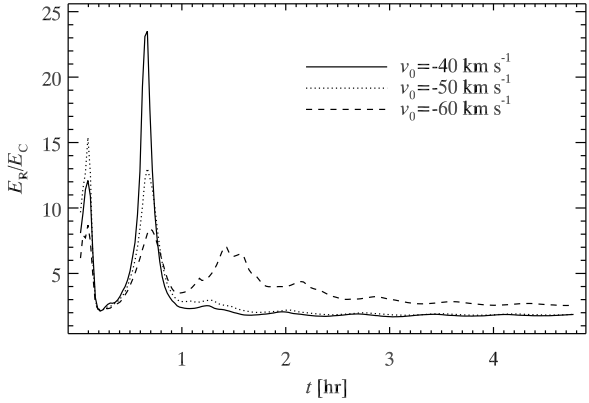}
\caption{Temporal variations of $E_{\rm R}/E_{\rm C}$ in the oscillation step
in the cases of $v_0=-40$, -50, and -60 km s$^{-1}$.}
\label{fig7}
\end{figure}

If the energy dissipation terms such as the radiative cooling and the heat
conduction are removed from Eq. (\ref{eqn3}), as we did in a test simulation,
we found that the prominence oscillation does not damp at all. When the two
non-adiabatic terms are kept, the prominence oscillation always damps. In order
to see the importance of the two terms, we calculate the time integrations of 
radiative loss ($E_{\rm R}$) and thermal conduction ($E_{\rm C}$) of the whole 
system after subtracting the corresponding values when the prominence is 
static at the center of the dip. Here $E_{\rm R}$ and $E_{\rm C}$ are the
integrals of the radiative and the conductive terms in the energy equation 
Eq. (\ref{eqn3}), where the integrals are taken in the whole corona above the
two footpoints. The evolutions of the ratio ($E_{\rm R}/E_{\rm C}$) in the
cases of $v_0=-40$, -50, and -60 km s$^{-1}$ are displayed in Fig.~\ref{fig7}.
It is seen that the ratio is always larger than unity. Especially in the early
stage of the oscillation when the amplitude is still large, $E_{\rm R}$ is even
one order of magnitude larger than $E_{\rm C}$. It is also revealed that as the
initial velocity increases, $E_{\rm R}$ becomes more and more important in
most of the lifetime of the oscillation. Our results support the 
conclusions of Terradas et al. (\cite{terr01,terr05}) that radiative loss is 
responsible for the damping of the slow mode of prominence oscillations in the
dip-shaped magnetic configurations, which seems to be different from the 
case of slow-mode waves propagating in the coronal loops where heat 
conduction contributes more to the damping (De Moortel et al. \cite{de02a,de02b}).

The role of the radiative cooling can be understood in a simple model as
follows: Since there are two segments of the corona in the magnetic loop, as
the prominence oscillates, one part would be attenuated and the other be
compressed. Suppose that the total length of the coronal part of the magnetic
loop is unity, which includes the part $x$, which is to the left of the
prominence, and the other part $1-x$, which is to the right of the prominence.
Hence, the densities of the corona on the two sides are proportional to $1/x$
and $1/(1-x)$, respectively. The total optically-thin radiative loss of the
coronal part is proportional to $x^{-2}+(1-x)^{-2}$, which is the minimum when
$x=0.5$, i.e., when the prominence is situated at the equilibrium position. 
Whenever the prominence deviates from the loop center, the cooling becomes
larger, dissipating the kinetic energy of the oscillating prominence. The model
is best illustrated by the relationship between the damping timescale ($\tau$)
and the initial amplitude of the oscillation, i.e., $A_0$ in Eq.
(\ref{eq-sin}). As $A_0$ increases, one of the two coronal parts is more
severely compressed, so the radiative cooling $x^{-2}+(1-x)^{-2}$ deviates
further away from the minimum value, i.e., it becomes larger. As a result, the
oscillation decays more rapidly.

Based on the sinusoidal function,  $A_0\propto v_0 P$. Substituting Eq. 
(\ref{eqn-new2}) into it, we get $A_0 \propto v_0 w D^{-1/2}$. With
this, it is not difficult to understand the positive correlation between the
decay timescale $\tau$ and $D$ and the negative correlation between $\tau$ and
$w$ as revealed by the lower row of Fig.~\ref{fig4}. Along this line of
thought, the dependence of the decay timescale on the prominence length can be
explained as follows: As the prominence thread is longer, the coronal part of
the magnetic loop, which radiates out the thermal energy, is shorter. More
importantly, the longer thread, with the same initial velocity, has a larger 
kinetic energy. Therefore, it takes a longer time for the compressed coronal
part to radiate it out.

\begin{figure}
\includegraphics[width=6cm]{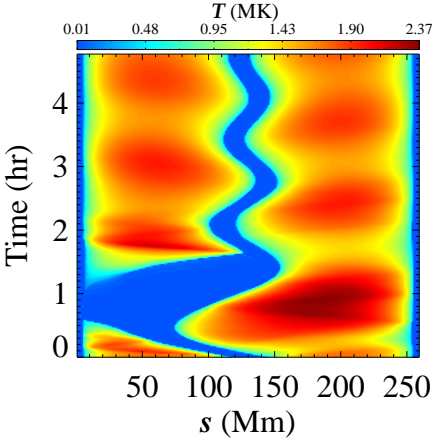}
\caption{Temporal evolution of the temperature along the magnetic loop when
the initial velocity perturbation is as large as $v_0=$-80 km s$^{-1}$. Note
that the prominence overpasses the magnetic loop apex and drains down to the
chromosphere at the left footpoint around $t=0.8$ hr.}
\label{fig8}
\end{figure}

It is seen from the first six cases (i.e., $|v_0|$ from 10 km 
s$^{-1}$ to 60 km s$^{-1}$) in the lower-right panel of Fig.~\ref{fig4} that
the decay timescale decreases with the initial perturbation velocity nearly
linearly. However, when $v_0$ is larger than 70 km s$^{-1}$, part of the
prominence would overpass the magnetic loop apex and drain down. The critical
velocity for the prominence to reach the loop apex can be roughly estimated as
$v_{criti}\sim \sqrt{2g_{\odot}D}=23\sqrt{D/{\rm Mm}}$ km s$^{-1}$.
Therefore, the value of $v_{criti}$ is 73 km s$^{-1}$ in the case of $D=$10 Mm.
As revealed from our simulations, even when $v_0$=-70 km s$^{-1}$, mass 
drainage already happens, although the amount of the drainage is much less 
than that in the case of $v_0=$-80 km s$^{-1}$. The temperature evolution along
the loop in the case of $v_0=$-80 km s$^{-1}$ is presented in Fig.~\ref{fig8}.
It is seen that part of the prominence falls down to the left leg of loop,
leading to the drainage of the prominence mass and kinetic energy as well,
while the remaining part continues to oscillate along the dip. The oscillation
period and the decay timescale in the cases with mass drainage are marked as
triangles in Fig.~\ref{fig4}. Their periods, $\sim$90.6 min, are slightly below
the trend defined by other cases without mass drainage ({\it diamonds}), which
is consistent with the weak positive correlation between $P$ and the prominence
length $l$. However, the damping timescales are greatly reduced, compared to
the trend defined by other cases without mass drainage as seen from the 
lower-right panel of Fig.~\ref{fig4}. Such a result, namely that mass drainage would
greatly reduce the decay timescale, might explain the mismatch between the
simulation and the observation of the decay of a prominence oscillation
reported in Zhang et al. (\cite{zhang12}).

\section{Summary} \label{S-summary}

In this paper, we carry out 1D hydrodynamic simulations of longitudinal
prominence oscillations using the MPI-AMRVAC code, extending 
earlier numerical simulations of prominence formation
(Xia et al. \cite{xia11}) and of prominence oscillations
(Luna \& Karpen \cite{luna12}; Zhang et al. \cite{zhang12}).
The simulations are divided into three steps:
First, a prominence forms and grows near the center of the dip-shaped
coronal loop due to chromospheric heating and the subsequent thermal
instability. Then, it relaxes to a quiescent state after the
chromospheric heating is switched off. Subjected to two kinds of 
perturbations that mimic subflares, the prominence starts to oscillate 
along the dip. 
Within the framework of the evaporation-condensation model, we obtained
scaling-laws for the prominence length ($l$) and mass ($M$), which are 
expressed as $l\sim \Delta t_1^{0.70}h^{-0.37}D^{-0.21}$ and $M\sim
\Delta t_1^{0.98}h^{-0.34}$, where $\Delta t_1$ is the time duration of
the chromospheric heating and evaporation, $h$ is the prominence height, $D$ is
the depth of the magnetic dip. It is found that $l$ is insensitive to the 
half length of the magnetic dip ($w$) once $w$ is large enough, say, 60 Mm; 
$M$ is insensitive to $D$ and $w$.
Both transient heating at one leg of the loop and an impulsive velocity
perturbation applied to the prominence as a whole are capable of driving a
coherent oscillation along the dip. The oscillation properties are 
found insensitive to the perturbation type in the regimes studied. 
In the case of the transient heating, $\sim$4\% of the deposited energy is 
converted into the kinetic energy of the prominence.
The longitudinal oscillations are sustained mainly by the tangential
component of gravity, except when the prominence is short and the gas
pressure gradient becomes also important. Both simulations and linear analysis
reveal that the period of oscillation ($P$) is 2$\pi\sqrt{R/g_\odot}$, where 
$R$ denotes the curvature radius of the dip, as also found by Luna \& Karpen
(\cite{luna12}). Other parameters, such as the length and the height of the 
prominence, as well as the perturbation velocity, also affect $P$, though
slightly. The longitudinal oscillations damp in the presence of 
non-adiabatic effects, i.e., radiative loss and thermal conduction 
(Soler et al. \cite{sol09}), among which the radiative
loss plays a leading role. With the parameter survey, we obtained a scaling-law
for the decay timescale $\tau$, which is expressed as $\tau\sim l^{1.63}
D^{0.66}w^{-1.21}v_{0}^{-0.30}$, where $v_0$ is the initial velocity
perturbation. We also found that prominence mass drainage, once it happens,
significantly reduces the decay timescale, which may explain the mismatching
between the simulations and the observations disclosed by Zhang et al.
(\cite{zhang12}). 

It is worth mentioning the limitation of the applications of the above
results. According to this paper, the mass of a prominence thread is
insensitive to the depth $D$ and the width $w$ of the magnetic dip. This is
based on the prominence formation directly via chromospheric evaporation with
a fixed lifetime $\Delta t_1$. According to Xia et al. (\cite{xia11}), the
prominence would grow via siphon flow even when the localized heating is
switched off, though the growth speed is much slower. Recently, Luna et al.
(\cite{ln12a}) pointed out that the restoring force of the longitudinal
oscillations depends on the depth of the magnetic dip. For shallow
dips, gas pressure plays an important role, while gravity is the main factor
for deep dips. Besides, Li \& Zhang (\cite{li12}) suggested that magnetic
tension may also contribute to the restoring force. As for the damping 
mechanisms, several other effects might be taken into account in the future
simulations, such as the wave leakage and plasma viscosity (Ofman \& Wang 
\cite{ofm02}). However, some will only be quantifiable in true multidimensional 
configurations, e.g. starting from the prominences formed in 
Xia et al. (\cite{xia11}).

\begin{acknowledgements}
The authors thank the anonymous referee for detailed and enlightening 
comments which improved the paper. Q. M. Zhang appreciates C. Fang, M. D.
Ding, W. Q. Gan, Y. P. Li, Z. J. Ning, S. M. Liu, D. J. Wu, H. Li, and L. Feng
for discussions and suggestions throughout this work. RK acknowledges 
funding from the Interuniversity Attraction Poles Programme initiated by 
the Belgian Science Policy Office (IAP P7/08 CHARM). The research is
supported by the Chinese foundations NSFC (11025314, 10878002, 10933003,
and 11173062) and 2011CB811402.
\end{acknowledgements}


\begin{thebibliography}{}
\bibitem[2000]{anti00} Antiochos, S.~K., MacNeice, P.~J., \& Spicer, D.~S.\
	2000, \apj, 536, 494
\bibitem[2010]{antolin10} Antolin, P., Shibata, K., \& Vissers, G.\ 2010, 
	\apj, 716, 154
\bibitem[2011]{arre11} Arregui, I., \& Ballester, J.~L.\ 2011, \ssr, 158, 169
\bibitem[2012]{arre12} Arregui, I., Oliver, R., \& Ballester, J.~L.\ 2012,
	Liv. Rev. Solar Phys., 9, 2
\bibitem[2006]{aul06} Aulanier, G., DeVore, C.~R., \& Antiochos, S.~K.\ 2006, 
        \apj, 646, 1349
\bibitem[2011a]{blok11a} Blokland, J.~W.~S., \& Keppens, R.\ 2011,
	\aap, 532, A94
\bibitem[2011b]{blok11b} Blokland, J.~W.~S., \& Keppens, R.\ 2011,
	\aap, 532, A93
\bibitem[2011]{bocc11} Bocchialini, K., Baudin, F., Koutchmy, S.,
	Pouget, G., \& Solomon, J.\ 2011, \aap, 533, A96
\bibitem[2011]{chen11} Chen, P.~F.\ 2011, Living Reviews
	in Solar Physics, 8, 1
\bibitem[2008]{chen08} Chen, P.~F., Innes, D.~E., \& Solanki, S.~K.\ 2008,
	\aap, 484, 487
\bibitem[2008]{colg08} Colgan, J., Abdallah, J., Jr., Sherrill, M.~E., et al.\
	2008, \apj, 689, 585
\bibitem[2002a]{de02a} De Moortel, I., 
        Ireland, J., Walsh, R.~W., \& Hood, A.~W.\ 2002, \solphys, 209, 61
\bibitem[2002b]{de02b} De Moortel, I., 
        Hood, A.~W., Ireland, J., \& Walsh, R.~W.\ 2002, \solphys, 209, 89	
\bibitem[2000]{dev00} DeVore, C.~R., \& Antiochos, S.~K.\ 2000, \apj, 539, 954
\bibitem[2002]{eto02} Eto, S., Isobe, H., 
        Narukage, N., et al.\ 2002, \pasj, 54, 481
\bibitem[2006]{fang06} Fang, C., Tang, Y.-H., \& Xu, Z.\ 2006, \cjaa, 6, 597
\bibitem[2010]{fang10} Fang, C., Chen, P.-F., Jiang, R.-L., \& Tang, Y.-H.\
	2010, Research in Astronomy and Astrophysics, 10, 83 
\bibitem[2008]{gil08} Gilbert, H.~R., Daou, A.~G., Young, D., 
        Tripathi, D., \& Alexander, D.\ 2008, \apj, 685, 629 
\bibitem[2010]{guo10} Guo, Y., Schmieder, B., D{\'e}moulin, P., Wiegelmann, T., 
        Aulanier, G., T{\"o}r{\"o}k, T., \& Bommier, V.\ 2010, \apj, 714, 343
\bibitem[2008]{han08} Hannah, I.~G., Christe, S., Krucker, S., et al.\ 2008, 
        \apj, 677, 704
\bibitem[2011]{her11} Hershaw, J., Foullon, C., Nakariakov, V.~M., \& Verwichte, E.\ 2011, 
        \aap, 531, A53
\bibitem[1986]{hood86} Hood, A.~W.\ 1986, \solphys, 105, 307
\bibitem[1966]{hyde66} Hyder, C.~L.\ 1966, \zap, 63, 78 
\bibitem[2006]{iso06} Isobe, H., \& Tripathi, D.\ 2006, \aap, 449, L17
\bibitem[2012]{jia12} Jiang, R.-L., Fang, C., \& Chen, P.-F.\ 2012, \apj, 751, 152
\bibitem[2006]{jing06} Jing, J., Lee, J., Spirock, T.~J., \& Wang, H.\ 2006, 
        \solphys, 236, 97
\bibitem[2003]{jing03} Jing, J., Lee, J., Spirock, T.~J., Xu, Y., Wang, H., 
        \& Choe, G.~S.\ 2003, \apjl, 584, L103
\bibitem[2008]{karp08} Karpen, J.~T., \& Antiochos, S.~K.\ 2008, \apj, 676, 658
\bibitem[2006]{karp06} Karpen, J.~T., Antiochos, S.~K., \& Klimchuk, J.~A.\ 2006, 
        \apj, 637, 531
\bibitem[2005]{karp05} Karpen, J.~T., Tanner, S.~E.~M., Antiochos, S.~K., 
        \& DeVore, C.~R.\ 2005, \apj, 635, 1319
\bibitem[2012]{kepp12} Keppens, R., Meliani, Z., van Marle, A.~J., et al.\ 2012, 
        Journal of Computational Physics, 231, 718
\bibitem[2003]{kepp03} Keppens, R., Nool, M., T{\'o}th, G., \& Goedbloed, J.~P.\ 2003, 
        Computer Physics Communications, 153, 317
\bibitem[1957]{kip57} Kippenhahn, R., \& Schl{\"u}ter, A.\ 1957, \zap, 43, 36
\bibitem[1969]{kle69} Kleczek, J., \& Kuperus, M.\ 1969, \solphys, 6, 72
\bibitem[1974]{kup74} Kuperus, M., \& Raadu, M.~A.\ 1974, \aap, 31, 189
\bibitem[2010]{lab10} Labrosse, N., Heinzel, P., Vial, J.-C. et al.\ 2010,
	\ssr, 151, 243
\bibitem[2012]{li12} Li, T., \& Zhang, J.\ 2012, \apjl, 760, L10 
\bibitem[2005]{lin05} Lin, Y., Engvold, O., Rouppe van der Voort, L., Wiik,
	J.~E.,\& Berger, T.~E.\ 2005, \solphys, 226, 239
\bibitem[2012]{luna12} Luna, M., \& Karpen, J.\ 2012, \apjl, 750, L1
\bibitem[2012a]{ln12a} Luna, M., D{\'{\i}}az, A.~J., \& Karpen, J.\ 
	2012a, \apj, 757, 98
\bibitem[2012b]{ln12b} Luna, M., Karpen, J.~T., \& DeVore, C.~R.\ 2012b, \apj,
	746, 30
\bibitem[2010]{mac10} Mackay, D.~H., Karpen, J.~T., Ballester, J.~L.,
	Schmieder, B., \& Aulanier, G.\ 2010, \ssr, 151, 333
\bibitem[2005]{men05} Mendoza-Brice{\~n}o, C.~A., Sigalotti, L.~D.~G., 
	\& Erd{\'e}lyi, R.\ 2005, \apj, 624, 1080
\bibitem[2004]{mul04} M{\"u}ller, D.~A.~N., Peter, H., \& Hansteen, V.~H.\ 
	2004, \aap, 424, 289 
\bibitem[2009]{ning09} Ning, Z., Cao, W., Okamoto, T.~J., Ichimoto, K.,
	\& Qu, Z.~Q.\ 2009, \aap, 499, 595
\bibitem[2002]{ofm02} Ofman, L., \& Wang, T.\ 2002, \apjl, 580, L85 
\bibitem[2007]{oka07} Okamoto, T.~J., Tsuneta, S., Berger, T.~E., et al.\ 2007, 
        Science, 318, 1577
\bibitem[2012]{schm12} Schmieder, B., \& Aulanier, G.\ 2012, EAS Publications
	Series, 55, 149
\bibitem[2002]{shim02} Shimizu, T., Shine, R.~A., Title, A.~M., Tarbell,
	T.~D., \& Frank, Z.\ 2002, \apj, 574, 1074
\bibitem[2009]{sol09} Soler, R., Oliver, R., \& Ballester, J.~L.\ 2009, \apj,
	693, 1601
\bibitem[2012]{su12} Su, Y., \& van Ballegooijen, A.\ 2012, \apj, 757, 168
\bibitem[1995]{tan95} Tandberg-Hanssen, E.\ 1995, Science, 269, 111
\bibitem[2001]{terr01} Terradas, J., Oliver, R., \& Ballester, J.~L.\ 
        2001, \aap, 378, 635
\bibitem[2005]{terr05} Terradas, J., Carbonell, M., Oliver, R., \& Ballester, J.~L.\ 2005, 
        \aap, 434, 741
\bibitem[2009]{tri09} Tripathi, D., Isobe, H., \& Jain, R.\ 2009, \ssr, 149, 283
\bibitem[1994]{van94} van der Linden, R.~A.~M., Hood, A.~W., \& Goedbloed, J.~P.\ 1994, 
        \solphys, 154, 69
\bibitem[2007]{vrn07} Vr{\v s}nak, B., Veronig, A.~M., Thalmann, J.~K., \& {\v Z}ic, T.\ 2007, 
        \aap, 471, 295
\bibitem[2012]{xia12} Xia, C., Chen, P.~F., \& Keppens, R.\ 2012, \apjl, 748, L26
\bibitem[2011]{xia11} Xia, C., Chen, P.~F., Keppens, R., \& van Marle, A.~J.\ 2011, 
        \apj, 737, 27
\bibitem[2012]{xu12} Xu, Z., Lagg, A., Solanki, S., \& Liu, Y.\ 2012, \apj, 749, 138
\bibitem[2012]{zhang12} Zhang, Q.~M., Chen, P.~F., Xia, C., \& Keppens, R.\ 2012, 
        \aap, 542, A52
\end{thebibliography}
\end{document}